\documentclass[10pt,superscriptaddress,tightenlines,twocolumn,amsmath,amssymb,aps, pra]{revtex4-2}
\UseRawInputEncoding
\usepackage[T1]{fontenc}
\usepackage[utf8]{inputenc}
\usepackage{lmodern}
\usepackage{CJKutf8}
\usepackage{mathrsfs}
\UseRawInputEncoding
\usepackage{graphicx}
\usepackage{dcolumn}
\usepackage{bbm}
\usepackage{color}
\usepackage{amsmath}
\usepackage{amssymb}
\usepackage{amsthm}
\usepackage{subfigure}
\usepackage{braket}
\usepackage{indentfirst}
\usepackage{mathrsfs}
\usepackage[linesnumbered,ruled,vlined]{algorithm2e}
\usepackage{enumerate}
\usepackage{enumitem}
\usepackage{algorithmic}
\usepackage{amsfonts}
\usepackage[normalem]{ulem} 

\usepackage{makecell}
\usepackage{lmodern}
\usepackage{lineno}
\usepackage{multirow}

\usepackage{newtxtext,newtxmath}

\usepackage{graphicx}
\usepackage{tabularx}
\usepackage{ulem}
\usepackage[backref=false,
           colorlinks=true,
           linkcolor=blue,
           citecolor=blue]{hyperref}
\usepackage{physics}
\usepackage{braket}
\usepackage{hyperref}
\usepackage{url}
\usepackage{booktabs}
\usepackage{multirow}
\usepackage{makecell} 

\newcommand{\be}{\begin{equation}}
\newcommand{\ee}{\end{equation}}

\newcommand{\RNum}[1]{\uppercase\expandafter{\romannumeral #1\relax}}

\makeatletter

\newcommand{\Rmnum}[1]{\expandafter\@slowromancap\romannumeral #1@}

\makeatother

\usepackage{xcolor}
\colorlet{RED}{red}
\colorlet{BLACK}{black}

\providecommand{\U}[1]{\protect\rule{.1in}{.1in}}

\hypersetup{colorlinks=true, linkcolor=red, citecolor=blue, urlcolor=black}

\makeatletter
\newcommand{\newreptheorem}[2]{\newtheorem*{rep@#1}{\rep@title}\newenvironment{rep#1}[1]{\def\rep@title{#2 \ref*{##1}}\begin{rep@#1}}{\end{rep@#1}}}
\makeatother

\newreptheorem{theorem}{Theorem}

\newreptheorem{lemma}{Lemma}

\newreptheorem{proposition}{Proposition}

\begin{document}

\title{Experimental demonstration of scalable quantum blockchain with exponentially superior quantum communication complexity}
\author{Feng Xie}
\thanks{These authors contributed equally to this work.}
\affiliation{School of Electrical Engineering, Guangxi University, Nanning 530004, China}
\affiliation{Guangxi Key Laboratory for Relativistic Astrophysics, School of Physical Science and Technology, Guangxi University, Nanning 530004, China}

\author{Ming-Yang Li}
\thanks{These authors contributed equally to this work.}
\affiliation{National Laboratory of Solid State Microstructures and School of Physics, Collaborative Innovation Center of Advanced Microstructures, Nanjing University, Nanjing 210093, China}
\affiliation{School of Physics and Key Laboratory of Quantum State Construction and Manipulation (Ministry of Education), Renmin University of China, Beijing 100872, China}

\author{Yongqiang Du}
\thanks{These authors contributed equally to this work.}
\affiliation{School of Electrical Engineering, Guangxi University, Nanning 530004, China}
\affiliation{Guangxi Key Laboratory for Relativistic Astrophysics, School of Physical Science and Technology, Guangxi University, Nanning 530004, China}

\author{Chen-Xun Weng}
\thanks{These authors contributed equally to this work.}
\affiliation{National Laboratory of Solid State Microstructures and School of Physics, Collaborative Innovation Center of Advanced Microstructures, Nanjing University, Nanjing 210093, China}
\affiliation{School of Physics and Key Laboratory of Quantum State Construction and Manipulation (Ministry of Education), Renmin University of China, Beijing 100872, China}

\author{Mingxuan Zhang}
\affiliation{School of Electrical Engineering, Guangxi University, Nanning 530004, China}
\affiliation{Guangxi Key Laboratory for Relativistic Astrophysics, School of Physical Science and Technology, Guangxi University, Nanning 530004, China}

\author{Xin Hua}
\affiliation{National Information Optoelectronics Innovation Center (NOEIC), Wuhan 430074, China}

\author{Xiang Guan}
\affiliation{School of Electrical Engineering, Guangxi University, Nanning 530004, China}
\affiliation{Guangxi Key Laboratory for Relativistic Astrophysics, School of Physical Science and Technology, Guangxi University, Nanning 530004, China}

\author{Xin An}
\affiliation{School of Electrical Engineering, Guangxi University, Nanning 530004, China}
\affiliation{Guangxi Key Laboratory for Relativistic Astrophysics, School of Physical Science and Technology, Guangxi University, Nanning 530004, China}

\author{Jingzhe~He}
\affiliation{Guangxi Key Laboratory of Multimedia Communications and Network Technology, School of Computer, Electronics, and Information, Guangxi University, Nanning 530004, China}

\author{Xin Liu}
\affiliation{School of Electrical Engineering, Guangxi University, Nanning 530004, China}
\affiliation{Guangxi Key Laboratory for Relativistic Astrophysics, School of Physical Science and Technology, Guangxi University, Nanning 530004, China}

\author{Zhenrong Zhang}
\affiliation{Guangxi Key Laboratory of Multimedia Communications and Network Technology, School of Computer, Electronics, and Information, Guangxi University, Nanning 530004, China}

\author{Xi Xiao}\email{xiaoxi@noeic.com}
\affiliation{National Information Optoelectronics Innovation Center (NOEIC), Wuhan 430074, China}

\author{Hua-Lei Yin}\email{hlyin@ruc.edu.cn}
\affiliation{School of Physics and Key Laboratory of Quantum State Construction and Manipulation (Ministry of Education), Renmin University of China, Beijing 100872, China}
\affiliation{National Laboratory of Solid State Microstructures and School of Physics, Collaborative Innovation Center of Advanced Microstructures, Nanjing University, Nanjing 210093, China}

\author{Kejin Wei}\email{kjwei@gxu.edu.cn}
\affiliation{School of Electrical Engineering, Guangxi University, Nanning 530004, China}
\affiliation{Guangxi Key Laboratory for Relativistic Astrophysics, School of Physical Science and Technology, Guangxi University, Nanning 530004, China}

\date{\today}
\begin{abstract}
To secure modern distributed digital infrastructures, quantum blockchains exploit quantum resources to achieve information-theoretic security and surpass the classical one-third fault-tolerance bound. However, existing high-fault-tolerant protocols face a fundamental scalability challenge: the blockchain trilemma imposes either exponential communication complexity or experimentally demanding multipartite entanglement. Here, we experimentally demonstrate a scalable quantum blockchain protocol based on weak coherent states that achieves an exponential reduction in quantum communication complexity. The protocol employs a circular quantum Byzantine agreement mechanism that preserves information-theoretic security while avoiding multipartite entanglement. We implement this protocol on a photonic integrated circuit platform, realizing a six-node network over commercially available telecommunication infrastructure. Compared with previous schemes, the protocol requires less than 4$\%$ of the quantum communication resources. Leveraging this advantage, we further demonstrate a quantum-secured token exchange application achieving a throughput of 805.3 transactions per second with zero failures. These results establish a practical pathway toward scalable quantum blockchain.
\end{abstract}

\maketitle

\section{Introduction}
Blockchain has emerged as a foundational information infrastructure for modern digital society, enabling decentralized data recording and trust establishment among mutually distrustful participants~\cite{swan2015blockchain}. As a distributed ledger technology, blockchain achieves immutability and consensus through cryptographic primitives and distributed agreement protocols, thereby eliminating reliance on centralized authorities. These properties position blockchain as a core enabler for applications ranging from digital currencies and decentralized finance to supply-chain management and distributed data governance~\cite{2019-Belotti,2025-Najmus}. Over the past decade, substantial progress has been made in improving classical blockchain performance-particularly in throughput, latency, and energy efficiency-while preserving decentralization and security under realistic adversarial models~\cite{2002-Castro,2019-yin-HBFT,2020-lu,2020-guo-dumbo}.

However, the security of existing blockchain systems fundamentally relies on classical computational assumptions that are increasingly challenged by advances in quantum computing~\cite{1994Shor,2018-fedorov,2019-arute,2020-Fernandez,2020-Zhong,2024-Acharya}. Widely deployed public-key cryptographic schemes, such as Rivest-Shamir-Adleman algorithm and elliptic-curve cryptography, are known to be vulnerable to polynomial-time quantum attacks, while quantum speedups also weaken the effective security of hash-based constructions. Beyond cryptographic primitives, classical blockchain consensus protocols are constrained by an intrinsic fault-tolerance limit: in classical decentralized networks, agreement cannot be achieved if more than one-third of the nodes behave adversarially~\cite{1986-dolev,1986-Fischer}. These limitations motivate the exploration of fundamentally new distributed ledger architectures that remain secure in the presence of quantum adversaries.

Quantum blockchains have therefore been proposed as a new paradigm that exploits quantum resources to achieve information-theoretic security and fault tolerance beyond classical limits~\cite{2024-Yang-BC}. By leveraging uniquely quantum features~\cite{wei2020high,2020-Xu,2023-Yin,li2025information,bozzio2025quantum}-such as the no-cloning theorem, measurement disturbance, and quantum correlations-quantum blockchain protocols can realize consensus and authentication mechanisms without computational hardness assumptions. In particular, quantum consensus models based on quantum Byzantine agreement (QBA) protocols can, in principle, tolerate higher fractions of dishonest participants than their classical counterparts, enabling more resilient decentralized systems and distributed applications in the quantum era~\cite{2001-Fitzi,iblisdir2004byzantine,2018-Kiktenko,2023-Weng,2025-Liu}.

Despite this promise, existing quantum blockchain proposals face substantial practical obstacles. High-fault-tolerance quantum consensus schemes are fundamentally constrained by a quantum analogue of the blockchain trilemma: achieving strong security and decentralization among $N$ participants typically results in either exponential communication complexity~\cite{2023-Weng,2024-Jing,2025-Lu} or requires heuristic multiparticle entanglement with special structures that lack rigorous proof~\cite{2001-Fitzi,2008-Gaertner,2008-Neigovzen,2015-Rahaman,rajan2019quantum,gao2020novel,yang2022decentralization}. These requirements severely limit scalability, as multipartite entanglement is fragile, resource-intensive, and difficult to maintain and distribute over realistic network infrastructures. As a result, most proposed quantum blockchain architectures remain far from experimental feasibility and large-scale deployment.

Here, we overcome these limitations by proposing and experimentally demonstrating a scalable quantum blockchain architecture based on weak coherent states that achieves an exponential reduction in quantum communication complexity. Our approach employs a circular QBA protocol~\cite{weng2026scalable} that avoids multipartite entanglement while preserving information-theoretic security and high fault tolerance. We implement this protocol experimentally using a photonic integrated circuit platform, realizing a six-node network comprising five user nodes and one certificate authority (CA) over commercially available telecommunication infrastructure. We achieve secure key generation rates exceeding 2.64~kbps and a consensus rate of 0.686 times per second (tps) for 1~Mbit messages. Moreover, our protocol consumes less than 4\% of the quantum communication resources required by the best-known quantum blockchain schemes, demonstrating a substantial and experimentally verified communication-complexity advantage.

Leveraging this exponential efficiency, we further demonstrate a practical application by implementing a quantum-secured token exchange system, a setting with many participants that is inaccessible to conventional quantum blockchain protocols. Our system supports end-to-end data acquisition, block recording, and token exchange, and achieves a sustained throughput up to 805.3 transactions per second with zero failed transactions, demonstrating feasibility beyond proof-of-principle implementations. Our results establish a realistic pathway toward scalable and deployable quantum blockchain systems, bridging quantum information theory and real-world distributed ledger applications.

\begin{figure*}[t] 
		\centering
		\includegraphics[width=1\textwidth]{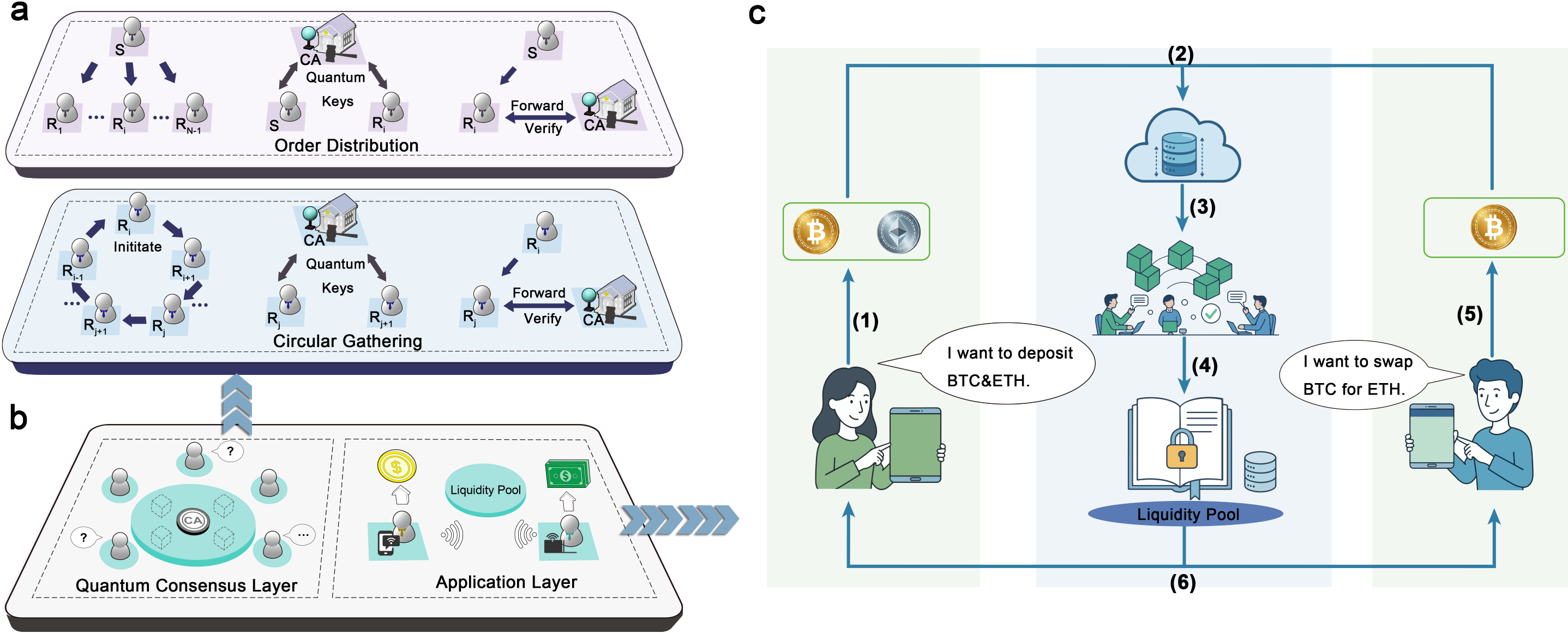}
		\caption{\textbf{ Quantum-consensus-enabled token exchange architecture and application.} 
            \textbf{a} Workflow of the circular QBA protocol for five users. This protocol involves a master node S and $N-1$ participant nodes R$_{1}$--R$_{N-1}$ as the users, with a CA acting as the verifier for every three-party QDS. The process executes in three phases: Order Distribution, Circular Gathering, and a final Consensus Output based on the collected authenticated messages.
			\textbf{b} Hierarchical architecture of the quantum-secure token exchange application. The architecture comprises two layers: the application layer, which handles decentralized-finance operations (e.g., token swaps and liquidity management) via smart contracts; and the quantum consensus layer, which ensures globally consistent transaction ordering using a circular QBA protocol. \textbf{c} Token exchange workflow built on the proposed architecture. \textbf{(1)} A liquidity provider (LP) initiates the cycle by depositing a token pair (e.g., Bitcoin (BTC) and Ether (ETH)) into the liquidity pool. \textbf{(2)} The backend service validates, formats, and submits the transaction to the network. \textbf{(3)} The ordering service utilizes the circular QBA consensus protocol to establish a tamper-resistant transaction sequence. \textbf{(4)} Validated blocks are committed to the distributed ledger, synchronizing the global state for reserves and ownership. \textbf{(5)} A Trader submits a swap request, which triggers the identical validation and consensus lifecycle (steps~2--4). \textbf{(6)} Users query the blockchain interface to verify pool balances and historical data, ensuring transparency without centralized intermediaries.}
		\label{network_architechure} 
	\end{figure*}

\section{Circular QBA protocol}
Our work employs a novel circular QBA protocol~\cite{weng2026scalable} designed to solve the Byzantine Generals Problem, in which distributed agents must reach a common decision in the presence of dishonest participants (Byzantine faults). The protocol leverages quantum digital signatures (QDS)~\cite{2023-Yin,du2025chip} to provide secure and non-repudiable communication among participants. It introduces a CA solely for signature verification; crucially, the CA does not participate in message generation, forwarding, or decision making. This restricted role yields a semi-decentralized architecture that simplifies network topology and reduces communication complexity.

Here, we provide a brief overview of the employed protocol. A detailed description of the protocol, together with analyses of its scalability and security, is provided in Appendix~\ref{app:qba}. As shown in Fig.~\ref{network_architechure}a, the protocol operates in three phases for a system of $N$ nodes, consisting of one randomly selected master node S and $N-1$ participant nodes R$_i$ ($i \in \{1,2,\cdots N-1\}$).

\textit{Order distribution:} The master node S distributes a signed message to each participant R$_i$ using one-time universal hashing (OTUH)-QDS with pre-shared quantum keys, where the CA acts as the verifier. Each participant receives a message that is information-theoretically secure against forgery and repudiation. If the master node is honest, all received orders are identical. If the master node is dishonest, different messages may be distributed; however, all signatures are recorded and verifiable by the CA.

\textit{Circular gathering: } Each participant independently initiates a clockwise message-gathering cycle involving all other participants. Authenticated message lists are iteratively appended and re-signed as they are forwarded. At each step, the CA verifies the validity and integrity of all signatures. If any signature is invalid, the process is restarted. Upon completion, all honest participants obtain an identical authenticated set of distributed messages.

\textit{Consensus output: }Each participant applies a predetermined deterministic function to the collected message list to produce its final output. Because all honest participants obtain identical authenticated data, consensus is guaranteed.

The security of the protocol follows directly from the information-theoretic non-repudiation and unforgeability of OTUH-QDS~\cite{2023-Yin}. The total failure probability is bounded by the cumulative forgery probability over all signing operations and can be made arbitrarily small by appropriate parameter choices. The protocol satisfies information-consistency conditions provided that at least two participants are honest, corresponding to a fault tolerance of $N > f+1$, where $f$ denotes the number of malicious parties.

Benefiting from its semi-decentralized structure, the protocol achieves a communication complexity of $\mathcal{O}(N^2)$, representing an exponential improvement over fully decentralized QBA schemes~\cite{2018-Kiktenko,2023-Weng,2024-Jing,2025-Lu}. This scalability makes the protocol well-suited for experimental implementation in real-world multi-user quantum networks.

\section{Token exchange application}
We develop a quantum-secure token exchange application built upon a layered architecture that integrates information-theoretically secure consensus with decentralized finance services. As shown in Fig.~\ref{network_architechure}b, the platform adopts a two-layer structure: a quantum consensus layer and an application layer. The quantum consensus layer is constructed based on the circular QBA protocol, which achieves distributed transaction ordering with $\mathcal{O}(N^{2})$ communication complexity, enabling scalable consensus among multiple participating nodes. On top of this foundation, the application layer hosts decentralized finance services implemented via smart contracts. 

The workflow of implementing a token exchange is illustrated in Fig.~\ref{network_architechure}c. The workflow involves two primary roles: liquidity providers (LPs) and traders. First, a liquidity provider deposits a token pair (BTC and ETH) into a liquidity pool (1). This action triggers a backend service request (2), which verifies the deposit, formats the transaction, and submits it to the blockchain network. Subsequently, the ordering service employs the circular QBA consensus protocol to reach distributed agreement on the global ordering of transaction batches (3).

Once consensus is reached, the validated blocks are synchronously committed to the distributed ledger (4). The ledger updates the global state by adjusting liquidity reserves or transferring asset ownership, while generating immutable records of liquidity provisioning and token exchange. Similarly, a trader initiates an exchange (e.g., swapping BTC for ETH) by submitting a transaction request (5), which undergoes the same verification and consensus procedures (steps 2-4). During this process, a smart contract executes the automated market maker pricing algorithm to complete the exchange, and the resulting state transitions are permanently recorded through the same commit mechanism.

Finally, both liquidity providers and traders can query the blockchain (6) to verify pool balances and historical records. The client interface presents this information to ensure transparency and auditability without reliance on centralized intermediaries. This end-to-end workflow demonstrates the system's ability to leverage a quantum-secure foundation to deliver a resilient decentralized exchange platform that is robust against both classical and quantum adversaries. 

Further details of the application, including system architecture, data interaction protocols between the frontend, backend, and blockchain network, are provided in Appendix~\ref{app:system}. The source code is publicly available at GitHub repository \url{(https://github.com/zmx-creator/Token-Exchange)}.

\begin{figure*}[t]
		\centering
		\includegraphics[width=1\textwidth]{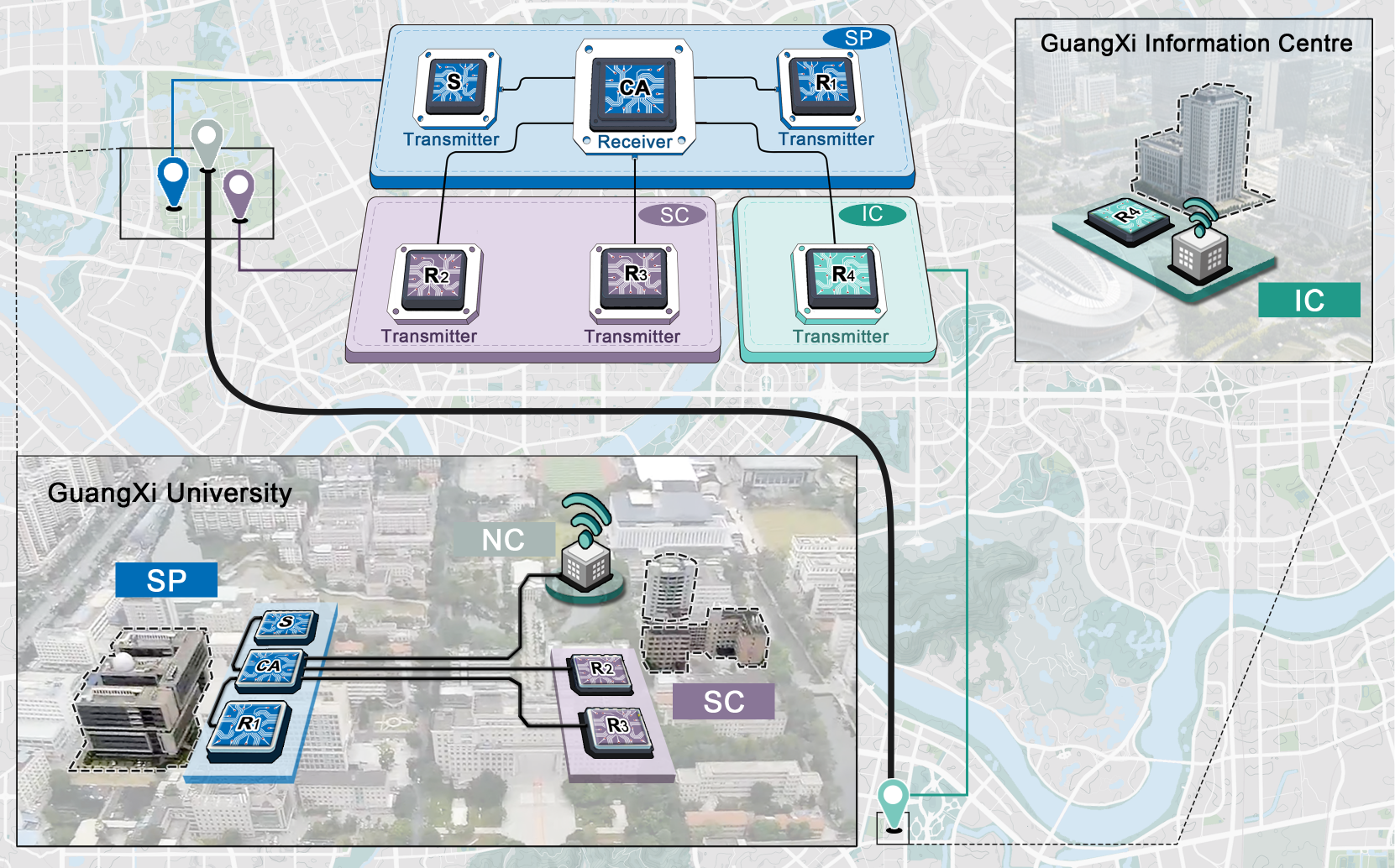}
		\caption{\textbf{Intercity deployment of scalable quantum blockchain network.}  A star-topology metropolitan quantum network implemented in Nanning, China, consisting of five user nodes (S, R$_1$-R$_4$) and a central CA. Two user nodes, S and R$_1$, and the certification authority (CA) are co-located in the School of Physical Science and Technology (SP). Two additional campus nodes, R$_2$ and R$_3$, are deployed in the School of Computer and Electronic Information (SC) and connected to the CA via an intermediate network. A remote user node is deployed at a metropolitan government information center (IC) is connected to the CA via the Guangxi University Network Center (NC) over commercial telecommunication fiber.}
		\label{application_schematic}
	\end{figure*}

\section{Field-deployed quantum network}

To experimentally validate the proposed circular QBA protocol with a token exchange application under realistic conditions, we deploy a field-scale quantum network operating over existing metropolitan fiber infrastructure. As illustrated in Fig.~\ref{application_schematic}, the network adopts a star topology centered on a CA, which naturally supports the semi-decentralized architecture of the protocol.

The deployed network consists of five user nodes and one CA node distributed across Nanning, China, spanning both campus and metropolitan environments. Two user nodes are co-located with the CA within the Guangxi University campus, while two additional campus nodes are connected via intermediate network facilities. A remote user node is deployed at a metropolitan government information center and connected through commercial telecom fiber links, resulting in end-to-end distances ranging from sub-kilometer to over 30~km. Across the five links, the network exhibits a mean link length of 7.1~km and an average channel loss of 12.0~dB, representative of realistic urban quantum communication conditions.

To enhance scalability and resource efficiency, all quantum receivers are centrally deployed at the CA, enabling multiple users to share bulky and costly detection hardware. Each user node is equipped with a compact quantum transmitter, forming a transmitter--receiver--shared architecture well suited for multi-user quantum networks. In this configuration, all user nodes establish information-theoretically secure keys with the CA using the one-decoy-state BB84 protocol~\cite{2018-Rusca}. These keys are subsequently consumed by the OTUH-QDS process employed in the circular QBA protocol, providing unforgeability and non-repudiation for all protocol messages.

Each user transmitter integrates a phase-randomized pulsed laser source, a silicon photonic encoder chip, a variable optical attenuator, and associated driving electronics, highlighting the compatibility of the protocol with photonic integrated platforms. The CA receiver incorporates a silicon photonic decoder chip with polarization tracking, multichannel superconducting nanowire single-photon detectors, and real-time control electronics, enabling stable operation across heterogeneous fiber links. Detailed point-to-point system configurations are provided in Appendix~\ref{app:qkd}.

\begin{figure*}[t] 
		\centering
\includegraphics[width=1\textwidth]{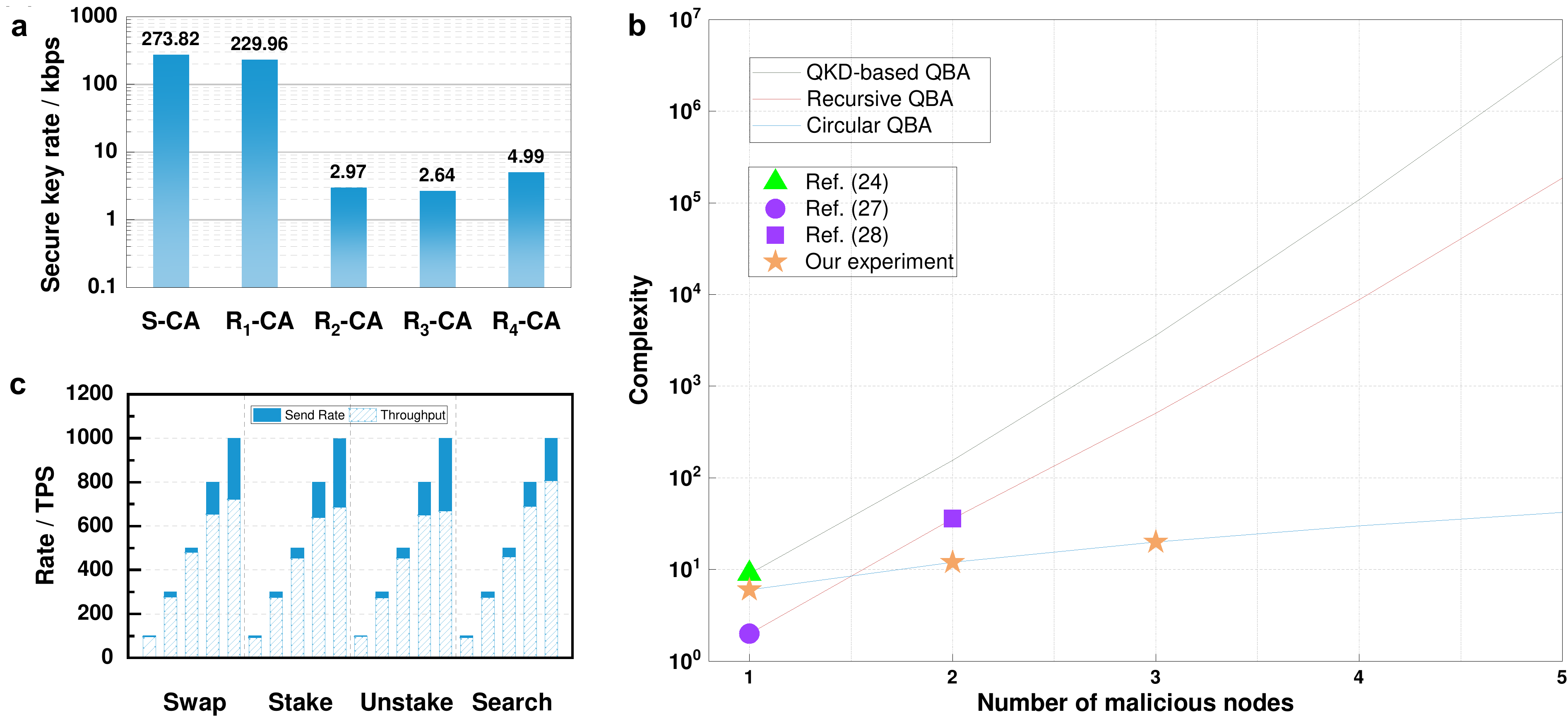}
		\caption{\textbf{ Performance of scalable quantum blockchain. } \textbf{a} The secure key rate across five links in the deployed quantum physical layer. The bar chart presents the measured secure key rates for the five point-to-point links in the network, with node pairs labeled on the horizontal axis and the corresponding secure key rates shown on the vertical axis. \textbf{b} Comparison of communication complexity. The green and red lines represent the exponential complexity of QKD-based~\cite{2018-Kiktenko} and recursive QBA~\cite{2023-Weng}, respectively, while the blue line shows the polynomial scaling of circular QBA. The green triangle marks the experiment of QKD-based QBA at $f=1$~\cite{2018-Kiktenko}, while the purple circle and square represent experimental demonstrations of recursive QBA for $f=1$~\cite{2024-Jing} and $f=2$~\cite{2025-Lu}, respectively. The yellow stars mark our experiments for $f=1, 2,$ and $3$ with circular QBA. \textbf{c} Stacked bar chart of transaction performance measured in transactions per second (TPS) using Hyperledger Caliper. Four transaction types are evaluated: Swap, asset exchange; Stake, liquidity provision via token locking; Unstake, liquidity withdrawal; and Search, ledger state and history queries. The upper segment denotes the send rate of submitted transactions, and the lower segment denotes the throughput of successfully committed transactions.}
		\label{result} 
	\end{figure*}

\begin{table*}[t]
\centering
\footnotesize
\caption{\textbf{Comparison of experimental quantum blockchain/consensus demonstrations} \textbf{$f_{\max}$}: maximum number of malicious nodes the system can tolerate; Communication Complexity: scaling of communication overhead with respect to the number of nodes; Experimental Complexity: the complexity for a single consensus round in the experiment; Chip-Based: using integrated photonic chip; Field Deployment: over real-world fiber; Entanglement-Free: without multi-partite entanglement; I.-T. Security: information-theoretic security; Introducing CA: employing a certificate authority.}
\label{tab:comparison}
\begin{tabular*}{\textwidth}{@{\extracolsep{\fill}} @{} ccccccccccc @{}}
\toprule
\textbf{} & 
\makecell{\textbf{Number}\\\textbf{of Users}} & 
\makecell{\textbf{$f_{\max}$}} & 
\makecell{\textbf{Communication}\\ \textbf{Complexity}} & 
\makecell{\textbf{Experimental}\\ \textbf{Complexity}} & 
\makecell{\textbf{Consensus}\\ \textbf{ Rate (tps)}} & 
\makecell{\textbf{Chip-} \\ \textbf{Based}} & 
\makecell{\textbf{Field}\\\textbf{Deployment}} & 
\makecell{\textbf{Entanglement-}\\\textbf{Free}} & 
\makecell{\textbf{I.-T.}\\\textbf{Security}} & 
\makecell{\textbf{Introducing}\\ \textbf{CA}} \\ 
\midrule

Ref.\cite{2008-Gaertner} & 3 & 1 & N/A & N/A & $1.6 \times 10 ^{-5}$ & $\times$ & $\times$ & $\times$ & $\times$ & $\times$ \\ 
\midrule

Ref.\cite{2018-Kiktenko} & 4 & 1 & \makecell{$\mathcal{O}(N^{f+1})$ \\ ($f < N/3$)} & 9 & 0.48 & $\times$ & \checkmark & \checkmark & \checkmark & $\times$ \\ 
\midrule

\multirow{2}{*}{Ref.\cite{2023-Weng}} & 5 & 2 & \multirow{2}{*}{\makecell{$\mathcal{O}(N^f)$ \\ ($f < N/2$)}}  & 36 & 0.54 & \multirow{2}{*}{$\times$} & \multirow{2}{*}{$\times$} & \multirow{2}{*}{\checkmark} & \multirow{2}{*}{\checkmark} & \multirow{2}{*}{$\times$} \\
 & 3 & 1 & & 2 & 11.95 & & & & & \\
\midrule

Ref.\cite{2024-Jing} & 3 & 1 & \makecell{$\mathcal{O}(N^f)$ \\ ($f < N/2$)} & 2 & $2.4 \times 10^{-3}$ & \checkmark & $\times$ & $\times$ & \checkmark & $\times$ \\ 
\midrule

Ref.\cite{2025-Lu} & 5 & 2 & \makecell{$\mathcal{O}(N^f)$ \\ ($f < N/2$)} & 36 & 0.057 & $\times$ & $\times$ & \checkmark & \checkmark & $\times$ \\ 
\midrule

\multirow{3}{*}{\textbf{This Work}} & 5 & 3 & \multirow{3}{*}{\makecell{$\mathcal{O}(N^2)$\\ ($f < N-1$)}}  & 20 & 0.686 & \multirow{3}{*}{\checkmark} & \multirow{3}{*}{\checkmark} & \multirow{3}{*}{\checkmark} & \multirow{3}{*}{\checkmark} & \multirow{3}{*}{\checkmark} \\
 & 4 & 2 & & 12 & 1.289 & & & & & \\
 & 3 & 1 & & 6 & 2.578 & & & & & \\ 
\bottomrule
\end{tabular*}
\end{table*}

\section{Experimental results}

We experimentally demonstrate a scalable quantum blockchain system implementing the circular QBA protocol over a field-deployed metropolitan quantum network. The experiments jointly evaluate the performance of the quantum physical layer, the quantum consensus layer, and the end-to-end blockchain application under realistic operating conditions.

At the quantum physical layer, five user nodes (S, R$_1$--R$_4$) establish information-theoretically secure keys with the  CA using the one-decoy-state BB84 protocol. The system parameters are globally optimized to accommodate heterogeneous fiber links spanning both campus-scale and metropolitan distances. For each user--CA link, we collect $10^{7}$ valid detection events as raw keys, followed by error correction and privacy amplification. The resulting secure key rates are 273.82, 229.96, 2.97, 2.64, and 4.99~kbps, respectively (Fig.~\ref{result}a), corresponding to a total secure key rate of approximately $(2.8$--$3.3)\times10^{6}$ bits per link. The detailed experimental results are provided in Appendix~\ref{app:exp-results}. These results confirm that stable and sufficient quantum key resources can be sustained across heterogeneous urban channels, providing a solid foundation for large-scale quantum consensus. 

We next evaluate the scalability and fault tolerance of the quantum consensus layer. The communication complexity of the circular QBA protocol, defined as the total number of required QDS operations, scales as
\begin{equation}
C(N)=N^{2}-N.
\end{equation}
Figure~\ref{result}b compares this scaling with existing high-fault-tolerance multi-party QBA protocols under identical adversarial assumptions. Owing to its fault-tolerance condition $N > f+1$, the circular QBA protocol remains secure provided that at least two participants are honest. Consequently, our five-node metropolitan network can tolerate up to three malicious nodes. To the best of our knowledge, this represents the first experimental quantum blockchain system achieving tolerance against three malicious nodes, whereas previous demonstrations were limited to one or two malicious nodes~\cite{2018-Kiktenko,2024-Jing,2025-Lu}. Notably, to tolerate three Byzantine faults ($f=3$), our protocol requires only 20 QDS rounds, whereas QKD-based QBA~\cite{2018-Kiktenko} and recursive QBA~\cite{2023-Weng} require 3,609 and 510 rounds, respectively. This corresponds to a reduction in communication complexity of approximately 96\% compared with recursive QBA at the same fault tolerance. When extrapolated to larger adversarial thresholds, the advantage increases to several orders of magnitude, demonstrating that the proposed architecture fundamentally overcomes the scalability bottleneck of QBA.

Beyond asymptotic complexity, we quantify the achievable consensus throughput. The consensus rate $R_{\mathrm{con}}$, defined as the lower bound on the number of successful consensus instances per second, is given by
\begin{equation}
R_{\mathrm{con}}=\frac{R_{\min}}{3L_{\mathrm{sig}}(N^{2}-N)},
\end{equation}
where $R_{\min}$ denotes the minimum secure key rate across all links and $L_{\mathrm{sig}}$ is the signature length. In the experiment, we set $L_{\mathrm{sig}}=64$, yielding a consensus-layer failure probability below $10^{-10}$. Under the five-party configuration ($N=5$), the system achieves a consensus rate of 0.686~tps with an overall failure probability of $\varepsilon_{\mathrm{con}}=10^{-11.76}$. For reduced configurations tolerating fewer malicious nodes, the consensus rate increases to 1.289~tps for $N=4$ and 2.578~tps for $N=3$, with correspondingly lower failure probabilities. Detailed calculations are provided in Appendix~\ref{app:qba}. 

Table~\ref{tab:comparison} summarizes these experimental results and compares them with previous demonstrations. The table shows that our experimental demonstration is the only implementation to date that combines integrated photonic chips, field deployment, and polynomial scalability. Unlike previous experiments that are either limited by the classical fault-tolerance bound~\cite{2018-Kiktenko} or the exponential complexity~\cite{2024-Jing,2025-Lu}, our fiver-user system achieves the highest experimental fault tolerance ($f_{\max}=3$) with significantly low complexity.

Finally, we assess the end-to-end system performance in a realistic blockchain application. Building on the QBA consensus layer, we deploy a token exchange workload on a six-node experimental setup, comprising five consensus participants and one CA node. The system is implemented in a VMware-based environment running Ubuntu~20.04 with 16~GB of RAM, and workload generation is performed using Hyperledger Caliper. As shown in Fig.~\ref{result}c, the system maintains a 100\% transaction success rate as the input load increases. The throughput scales approximately linearly up to 805.3 transactions per second before saturating (detailed benchmarking metrics are provided in Appendix~\ref{app:caliper}), demonstrating that quantum-secured consensus can support practical, high-throughput blockchain workloads in real-world network environments.

\section{Conclusion and discussion}
We have experimentally demonstrated a scalable quantum blockchain architecture operating over a real-world metropolitan fiber network, integrating an information-theoretically secure quantum physical layer with a quantum consensus layer implemented using photonic integrated circuits. By introducing a circular QBA protocol based on weak coherent states, our approach eliminates the need for multipartite entanglement and overcomes the exponential communication overhead that has long constrained high-fault-tolerance quantum blockchains. As a result, the communication complexity is reduced to a polynomial scaling in the number of participants, enabling practical multi-user deployment.

In a five-node metropolitan prototype network, the system operates under realistic channel conditions and achieves secure key generation rates exceeding 2.64~kbps per link, together with a consensus throughput of 0.686~tps for megabit-scale messages. Crucially, the system remains robust in the presence of up to three malicious nodes, representing the highest Byzantine fault tolerance achieved to date in an experimental quantum blockchain. Beyond consensus, we further demonstrate a quantum-secured token exchange application that supports asset transfer, ledger recording, and state querying, highlighting the feasibility of quantum blockchain technologies for trusted digital asset management and distributed applications.

More broadly, this work establishes system-level scalability across the quantum physical, consensus, and application layers, providing a concrete experimental foundation for future large-scale quantum blockchain networks. Looking ahead, further expansion in network size, deeper integration of photonic quantum hardware, and the exploration of application-specific quantum blockchain services are expected to accelerate the transition of quantum blockchains from laboratory demonstrations to practical components of next-generation secure digital infrastructure.

\bigskip
\noindent	
\begin{center}
    \textbf{\large Acknowledgements}
\end{center}
We thank Pingping Li for drawing the figures. This study was supported by Quantum Science and Technology-National Science and Technology Major Project (Grant No. 2025ZD0300200), the National Natural Science Foundation of China (Grants No.~12522419, No.~U25D8016, No.~12274223, No.~62171144, No.~62031024, and No.~62171485), Guangxi Science Foundation (Nos. 2025GXNSFAA069137, 2026GXNSFHA00640303), Guangdong Basic and Applied Basic Research Foundation (2024B1515120030), and Bagui Scholars Programme (W.X.-G., GXR-6BG2424001, GXR-1BGQ2424005).

\appendix
\setcounter{figure}{0}
\setcounter{table}{0}
\renewcommand{\thefigure}{S\arabic{figure}}
\renewcommand{\thetable}{S\arabic{table}}
\renewcommand{\theHfigure}{S\arabic{figure}}
\renewcommand{\theHtable}{S\arabic{table}}
\section{Details of Circular QBA Protocol}\label{app:qba}

\subsection{OTUH-QDS}

We first detail the one-time universal$_2$ hashing quantum digital signature (OTUH-QDS) protocol from Ref.~\cite{2023-Yin}, suitable for our circular QBA due to its efficient multi-bit message signing capability. This protocol comprises distribution and messaging stages.

In the distribution stage, the signer, forwarder, and verifier share keys $\{X_s, Y_s, Z_s\}$, $\{X_f, Y_f, Z_f\}$, and $\{X_v, Y_v, Z_v\}$, respectively, satisfying $X_v = X_s \oplus X_f$, $Y_v = Y_s \oplus Y_f$, and $Z_v = Z_s \oplus Z_f$. These keys (length $n$ bits each) are distributed via quantum secret sharing or quantum key distribution; the message ($M$) length is $m$ bits.

During the messaging stage, the signer generates a random irreducible polynomial over GF(2), representing its coefficients (excluding the leading coefficient) as an $n$-bit string $p$. Using its key $X_s$ and $p$, the signer constructs an LFSR-based Toeplitz hash function $h_{p,s}$, computes the $n$-bit hash $h_{p,s}(M)$ of the $m$-bit message $M$, and generates the signature $\text{Sig} = h_{p,s}(M) \oplus Y_s$. The signer then sends $\{M, \text{Sig}, p_s = p \oplus Z_s\}$ to the forwarder. The forwarder forwards $\{M, \text{Sig}, p_s\}$ to the verifier, with its keys $\{X_f, Y_f, Z_f\}$. Using the received keys and its own keys, the verifier reconstructs $\{X_s, Y_s, Z_s\}$, decrypts $p$ using $Z_s$, recover $h_{p,s}(M)$ from $\text{Sig}$ using $Y_s$, and recomputes a hash value $h^{\prime}_{p,s}(M)$ using $X_s$ and $p$. If two hash values match, the verifier accepts the signature and sends this result and its keys $\{X_v, Y_v, Z_v\}$ to the forwarder. The forwarder performs a similar verification process.

A secure QDS protocol must guarantee non-repudiation (the signer cannot deny authorship) and unforgeability (signatures cannot be forged). As shown in Ref.~\cite{2023-Yin}, the key correlation in OTUH-QDS ensures that the forwarder and verifier reach the same decision, making repudiation impossible, except for the negligible failure probability of secure message authentication. On the other hand, a forgery attempt requires guessing the irreducible polynomial used to generate the signature, rather than directly guessing the key. Using the properties of universal$_2$ hashing with LFSR-based Toeplitz matrices~\cite{1994-krawczyk}, the probability of successful forgery is bounded by $\varepsilon_{\rm{for}}(m) = \frac{m}{2^{L_{\mathrm{sig}}-1}}.$

\subsection{N-Party Circular QBA}

\begin{figure*}
    \centering
    \includegraphics[width=0.8\textwidth]{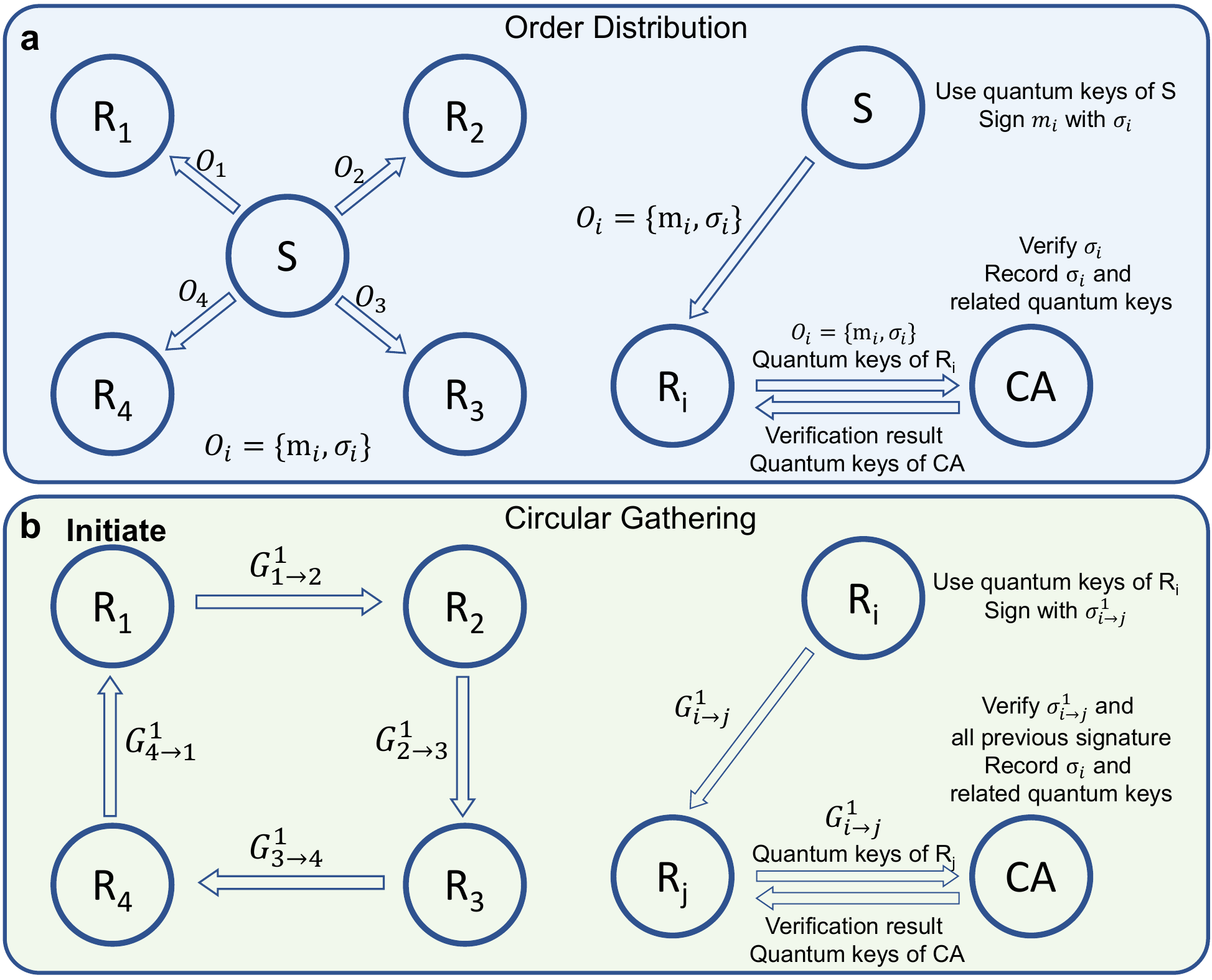}
    \caption{\textbf{Schematic of the circular QBA protocol.} The diagram illustrates the execution flow in our five-node experiment with S acting as the master node and R$_1$--R$_4$ as backup nodes. \textbf{(a) Order Distribution:} S generates initial orders $O_i=\{m_i, \sigma_i\}$ and distributes them to each backup node via three-party OTUH-QDS. In each transaction, the CA uses shared quantum keys to verify the signature $\sigma_i$ and records it for future verification. \textbf{(b) Circular Gathering:} Each backup node initiates message collection cycles. The diagram details the circular gathering initiated by R$_1$: R$_1$ signs its received message $O_1$ to create package $G_{1\to 2}^1$ and sends it to R$_2$. R$_2$ appends its own order $O_2$, signs the combined package, and sends $G_{2\to 3}^1$ to R$_3$. This clockwise relay continues until the package returns to R$_1$. At every step, the CA acts as the verifier, checking the validity of the new signature and the integrity of all previous signatures.}
    \label{fig:protocol}
\end{figure*}

Illustrated in Fig.~\ref{fig:protocol}, the circular QBA protocol~\cite{weng2026scalable} comprises three main stages: order distribution, circular gathering, and consensus output. We detail these steps below.

\begin{figure*}
	\centering
	\includegraphics[width=0.85\textwidth]{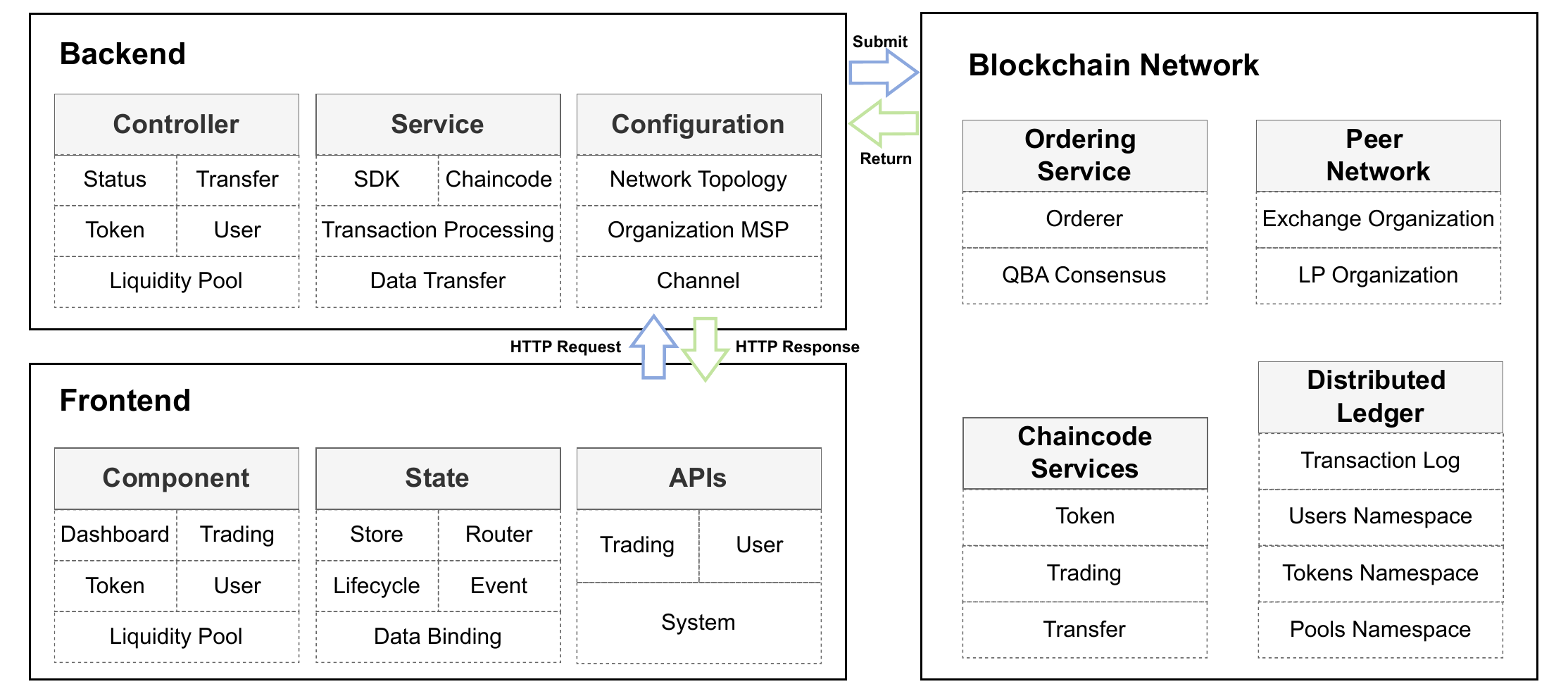}
	\caption{\textbf{System architecture and data interaction of the token exchange application.} The application comprises three layers-Frontend, Backend, and Blockchain Network-connected by bidirectional data flow. User actions in the Frontend are sent via HTTP to the Backend, where SDK-mediated transactions are submitted to the Blockchain Network. After chaincode execution and consensus-based ledger updates, the verification results and state changes are propagated back to the Frontend through the Backend. \textbf{Frontend:} Provides a user interface with modules for dashboard monitoring, trading, token and user administration, and liquidity pool management. These are synchronized by a state module, which manages application state via store, router, lifecycle, event, and data binding, communicating with the backend through dedicated HTTP APIs for trading, user, and system services. \textbf{Backend}: Handles business logic and blockchain interaction. The Controller module offers endpoints for network status, asset transfer, token and user operations, and liquidity pool updates. Services use the Fabric SDK to propose transactions, invoke chaincode, collect endorsements, and submit them to the ordering service, implementing transaction processing and data transfer. The configuration module defines critical parameters such as network topology, organization MSP identities, and channel profiles. \textbf{Blockchain Network}: Provides distributed trust via an ordering service. Orderer nodes achieve QBA consensus with quantum keys, and a peer network comprising exchange and LP organizations with endorsing peers. Chaincode services enforce on-chain logic for token, trading, and transfer. The distributed ledger stores transaction logs and world states in isolated namespaces for users, tokens, and pools.}
    \label{data_interaction}
\end{figure*}

1. Order distribution: The master node S generates an initial message $m_i$ and a signature $\sigma_i$ for each backup node R$_i$ ($i \in \mathbb{Z}_{N-1}^+$). This package $\{m_i;\sigma_i\}$ is sent to R$_i$ using a three-party OTUH-QDS, which involves S (signer), R$_i$ (forwarder), and the CA (verifier). After the signature is verified, R$_i$ records its initial message list as $O^{i} = \{m_i;\sigma_i\}$. If S is honest, all distributed messages are identical, satisfying $m_i=m_0~(\forall i\in \mathbb{Z}_{N-1}^+)$.

2. Circular gathering: Each of the $N-1$ backup nodes starts a message collection cycle that moves clockwise through $N-1$ steps. Taking the cycle started by R$_i$ as an example, the transmission follows the sequence $(i \rightarrow i+1\rightarrow \cdots\rightarrow N-1 \rightarrow 1 \rightarrow \cdots \rightarrow i)$.

In the first step, R$_i$ signs its initial list $O^i$ with $\sigma^i_{i\rightarrow i+1}$ and sends the package $G^i_{i\rightarrow i+1}=\{m_i;\sigma_i;\sigma^i_{i\rightarrow i+1}\}$ to R$_{i+1}$. In a general step within the cycle, a backup node R$_j$ ($j\neq i$) receives the package $G^i_{j-1\rightarrow j}$ from the previous node R$_{j-1}$. R$_j$ then appends its own initial message list $O^{j}=\{m_j;\sigma_j\}$ to the received data, signs the combined set with $\sigma^i_{j\rightarrow j+1}$, and forwards the updated package $G^i_{j\rightarrow j+1}$ to R$_{j+1}$. This is expressed as $G^i_{j\rightarrow j+1}= G^i_{j-1\rightarrow j}  \cup O^j \cup \{\sigma^i_{j\rightarrow j+1}\} = \{m_i,\cdots,m_j;\sigma_i,\cdots,\sigma_j;\sigma^i_{i\rightarrow i+1},\cdots,\sigma^i_{j\rightarrow j+1} \}$. These transfers are also performed using three-party OTUH-QDS.

When this round of circular gathering is finished, R$_i$ obtains the final list $G^i_{i-1\to i}$, which contains three parts: (1) the set of orders received by all backup nodes $\{m_1,m_2,\dots,m_{N-1}\}$; (2) the corresponding signatures from the master node $\{\sigma_1,\sigma_{2},\dots,\sigma_{N-1}\}$; and (3) the sequence of signatures generated by backup nodes during this round $\{\sigma^i_{i\to i+1},\sigma^i_{i+1\to i+2},\dots,\sigma^i_{N-1\to 1},\dots,\sigma^i_{i-1\to i}\}$.

We note that the CA records the signatures and the corresponding quantum keys in all QDS processes, and the CA checks the information package at every step in the circular gathering. First, it verifies the validity of the current signature $\sigma^i_{j\rightarrow j+1}$ as standard QDS. Second, it confirms that all previous signatures in the packet ($\{\sigma_i,\cdots,\sigma_j\}$ and $\{\sigma^i_{i\rightarrow i+1},\cdots,\sigma^i_{j-1\rightarrow j}\}$) remain valid and unchanged compared to the records. Since the CA stores the correlated keys and signatures from all previous OTUH-QDS sessions, this check is straightforward. If any verification fails, the gathering round initiated by R$_i$ is aborted and restarted.

3. Consensus output: After the circular gathering stage, every backup node R$_i$ $(\forall i \in \mathbb{Z}_{N-1}^+)$ holds an identical final message set $F^i = \{ m_1, m_2, \dots, m_{N-1} \}$. They apply a predefined deterministic function $D$ to this set to get the final decision $m^{N_i}=D(m_1, m_2, \dots, m_{N-1})$. Since $D$ is deterministic, identical inputs will result in the same output for all honest nodes.

We then analyze the performance of this consensus mechanism from four critical dimensions regarding security and scalability.

Fault-tolerance: The protocol surpasses the classical one-third fault-tolerance bound. When the master node is honest, unforgeability ensures all backup nodes receive the correct order. Because of the verification of CA, all backup nodes can not send any fake messages in the gathering phase. When the commander is dishonest, the CA-assisted verification also ensures all lieutenants gather an identical set of conflicting messages, leading to a deterministic default output. This mechanism requires only two honest participants (excluding the CA) to function, achieving a fault tolerance of
\begin{equation}
    N  > f + 1.
\end{equation}

Information-theoretical security: The security of the protocol is guaranteed by the unforgeability of the underlying OTUH-QDS, where the forgery probability $\varepsilon_{\rm{for}}(m) = \frac{m}{2^{L_{\mathrm{sig}}-1}}$ depends on the message length $m$ and the length of signature $L_{\mathrm{sig}}$. Since the message packet grows during the circular gathering phase, we strictly bound the failure probability using the maximum possible message length $L_{\max} = (N-1)m + (2N-3)L_{\mathrm{sig}}$. By analyzing the cumulative forgery risks under both honest and dishonest master node scenarios, the total security failure probability is given by:
\begin{equation}
	\begin{aligned}
		\varepsilon_{\mathrm{QBA}} = \max \Big\{ & f_{\max} \left[ \varepsilon_{\mathrm{for}}(m) + (N-f_{\max}-1) \varepsilon_{\mathrm{for}}(L_{\max}) \right], \\
		& (f_{\max}-1)(N-f) \varepsilon_{\mathrm{for}}(L_{\max}) \Big\},
	\end{aligned}
\end{equation}
where $m$ is the initial message length, and $f_{\max}=N-2$ is the maximum number of malicious nodes tolerated.

Communication complexity: This is the protocol's most significant advantage. The communication complexity $C(N)$, defined as the minimum number of QDS operations required, scales quadratically rather than exponentially:
\begin{equation}
   C(N)=(N-1)+(N-1)(N-1)=N^2-N,
\end{equation}
since there are $N-1$ QDS in the distribution phase, and $N-1$ rounds of circular gathering with each consisting of $N-1$ QDS. Our protocol demonstrates polynomial behavior, offering a distinct scalability advantage over the exponential growth characteristic of prior works~\cite{2018-Kiktenko,2023-Weng}.

Number of quantum channels: In the circular QBA, all the message distribution and transmission are completed using three-party OTUH-QDS, which requires three participants to share correlated quantum keys. Specifically, the key of one participant is the bitwise XOR sum of the keys held by the other two. As the CA functions as the verifier in all QDS, it can generate quantum keys with the other two participants independently, and calculate the XOR sum locally as its quantum key used in OTUH-QDS. This enables the circular QBA to be performed on a star-topology quantum network, where the user nodes only have a quantum channel connecting to the CA. Consequently, this architecture simplifies the network structure, reducing the number of quantum channels from the $N(N-1)/2$ required by a fully connected network~\cite{2018-Kiktenko,2023-Weng} to just $N$.

\begin{figure*}
	\centering
	\includegraphics[width=0.85\textwidth]{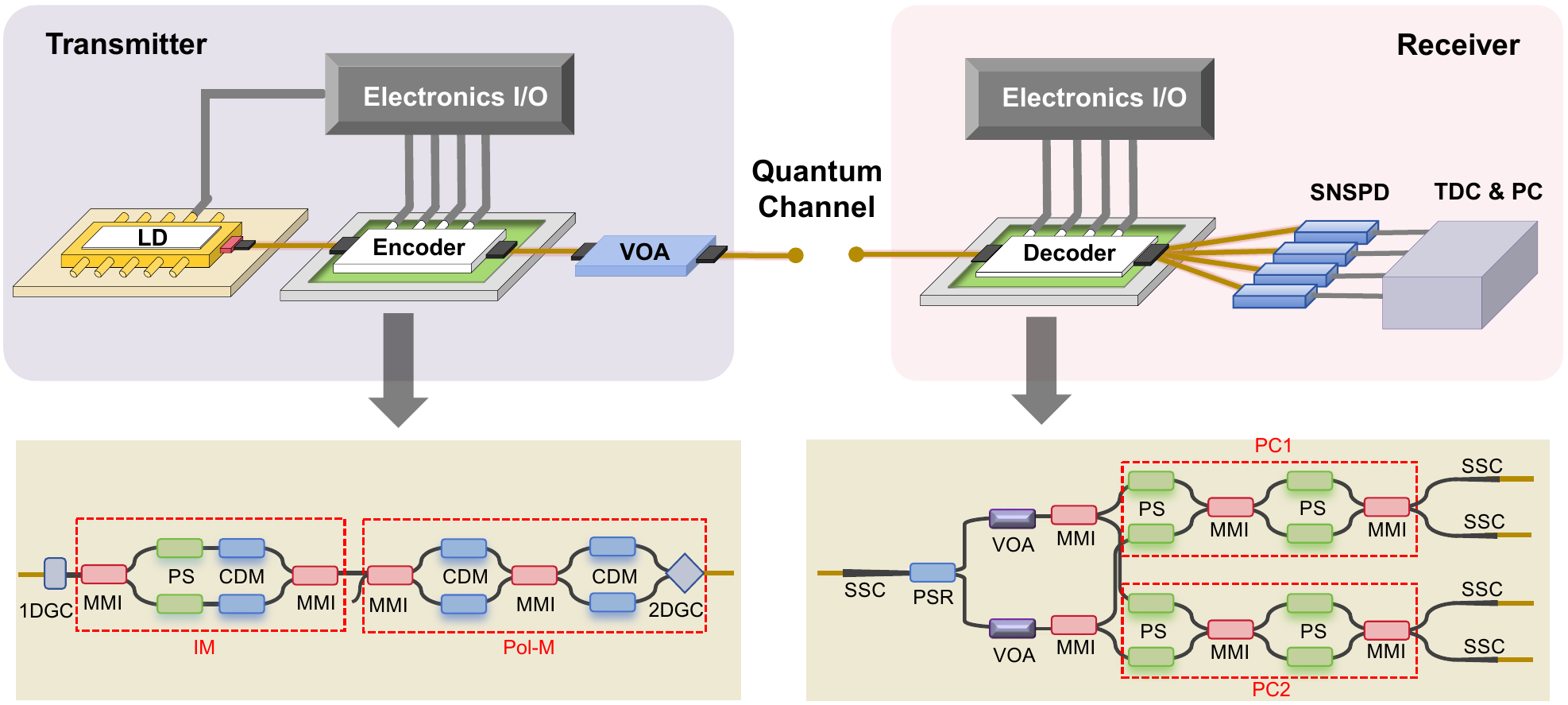}
	\caption{\textbf{Chip-integrated experimental setup for point-to-point quantum links. } The integrated quantum signal transmitter (left) generates polarization-encoded quantum states and transmits them to the integrated quantum signal receiver (right) through the quantum channel, in which quantum state analysis and single-photon detection are performed. The transmitter consists of a phase-randomized laser diode (LD), a silicon-based encoder chip (Encoder) dedicated to implementing decoy-state and polarization state preparation, and an off-chip variable optical attenuator (VOA). The receiver comprises a silicon-based decoder chip (Decoder) with polarization tracking capability, a multichannel superconducting nanowire single-photon detector (SNSPD), a time-to-digital converter (TDC), and a personal computer (PC) used for data acquisition and post-processing. The silicon-based encoder chip integrates intensity and polarization modulators (IM and Pol-M) realized using multimode interferometers (MMIs), thermo-optic modulators (TOMs), and carrier-depletion modulators (CDMs). One-dimensional and two-dimensional grating couplers (1DGC and 2DGC) are employed for efficient fiber-to-chip and chip-to-fiber optical coupling. The silicon-based decoder chip integrates a spot-size converter (SSC) for optical coupling, a polarization splitter-rotator (PSR) that maps polarization information onto paths, variable optical attenuators (VOAs) for compensating polarization-dependent loss of the PSR, and polarization controllers (PC1 and PC2) realized using MMIs and TOMs, thereby enabling high-precision quantum state analysis and polarization tracking.}
    \label{Chip_setup}
\end{figure*}

\begin{table*}[t]
	\centering
	\caption{\textbf{Experimental parameters and results.}Loss denotes the fiber transmission loss between each of the five user nodes (S, R$_1$-R$_4$) and the CA node. $\mu$ ($\nu$) denotes the mean photon number of the signal (decoy) pulses emitted by the transmitter, and $P_{\mu}$ ($P_{\nu}$) denotes the probability of randomly selecting the signal (decoy) state for state preparation. $P_{Z}$ ($P_{X}$) denotes the probability that the transmitter prepares quantum states in the $Z$ ($X$) basis. $n_{Z,\mu}$ ($n_{Z,\nu}$) denotes the total number of detection events recorded when both parties select the $Z$ basis and the transmitter emits signal (decoy) states, whereas $m_{Z,\mu}$ ($m_{Z,\nu}$) denotes the corresponding number of erroneous detection events. Similarly, $n_{X,\mu}$ ($n_{X,\nu}$) denotes the total number of detection events recorded when both parties select the $X$ basis and the transmitter emits signal (decoy) states, and $m_{X,\mu}$ ($m_{X,\nu}$) denotes the corresponding number of erroneous detection events. $n_{Z}$ denotes the total number of detection events in the $Z$ basis after basis sifting, and $t$ denotes the time required to accumulate $n_{Z}$ events. $s_{Z,1}^{l}$ denotes the lower-bound estimate of the number of single-photon detection events in the $Z$ basis. $E_{Z}$ denotes the quantum bit error rate in the $Z$ basis, and $\phi_{Z}^{u}$ denotes the upper bound on the phase error rate for single-photon events in the $Z$ basis. $\lambda_{\rm EC}$ denotes the number of information bits disclosed during error correction. $\mathrm{SKR}$ denotes the final secure key rate achievable for the corresponding link.}
	\label{table_rawdate}
	 \begingroup
		\fontsize{10}{12}\selectfont
		\setlength{\tabcolsep}{5.5pt}
		\renewcommand{\arraystretch}{1.2}

		\resizebox{\textwidth}{!}{%
		\begin{tabular}{*{12}{c}}
	\hline\hline
	\rule[-1ex]{0pt}{3.5ex}Link & Loss(dB) & $\mu$ & $\nu$ & $P_{\mu}$(\%) & $P_{\nu}$(\%) & $P_{Z}$(\%) & $P_{X}$(\%) & $n_{Z,\mu}$ & $m_{Z,\mu}$ & $n_{X,\mu}$ & $m_{X,\mu}$  \\ \hline
	\rule[-1ex]{0pt}{3.5ex}S-CA & 0.7 & 0.54 & 0.15 & 77.91 & 22.09 & 94.01 & 5.99 & 9287141 & 70348 & 475151 & 6345  \\ \hline
	\rule[-1ex]{0pt}{3.5ex}R$_1$-CA & 1.0 & 0.54 & 0.15 & 77.86 & 22.14 & 93.61 & 6.39 & 9326675 & 97210 & 487483 & 3575  \\ \hline
	\rule[-1ex]{0pt}{3.5ex}R$_2$-CA & 20.0 & 0.53 & 0.14 & 77.70 & 22.30 & 94.35 & 5.65 & 9290631 & 134860 & 449467 & 4013  \\ \hline
	\rule[-1ex]{0pt}{3.5ex}R$_3$-CA & 20.6 & 0.53 & 0.14 & 77.64 & 22.36 & 94.35 & 5.65 & 9286205 & 119319 & 463004 & 5028  \\ \hline
	\rule[-1ex]{0pt}{3.5ex}R$_4$-CA & 17.9 & 0.53 & 0.14 & 77.73 & 22.27 & 93.61 & 6.39 & 9283835 & 134610 & 416713 & 3315  \\ \hline\hline
	\\ \hline\hline
	\rule[-1ex]{0pt}{3.5ex}Link & $n_{Z,\nu}$ & $m_{Z,\nu}$ & $n_{X,\nu}$ & $m_{X,\nu}$ & $n_{Z}$ & $t$(s) & $s_{Z,1}^{l}$ & $E_{Z}$(\%) & $\phi_{Z}^{u}$(\%) & $\lambda_{EC}$ & SKR(kbps)  \\ \hline
	\rule[-1ex]{0pt}{3.5ex}S-CA & 712859 & 4555 & 37042 & 541 & $10^7$ & 12.1 & 5433605 & 0.75 & 4.10 & 768603 & 273.82  \\ \hline
	\rule[-1ex]{0pt}{3.5ex}R$_1$-CA & 673325 & 11423 & 36070 & 341 & $10^7$ & 12.3 & 4688184 & 1.09 & 2.56 & 1048140 & 229.96  \\ \hline
	\rule[-1ex]{0pt}{3.5ex}R$_2$-CA & 709369 & 12436 & 35289 & 386 & $10^7$ & 992.7 & 5263512 & 1.47 & 2.68 & 1377162 & 2.97  \\ \hline
	\rule[-1ex]{0pt}{3.5ex}R$_3$-CA & 713795 & 7885 & 36753 & 423 & $10^7$ & 1155.8 & 5371856 & 1.27 & 3.25 & 1214939 & 2.64  \\ \hline
		\rule[-1ex]{0pt}{3.5ex}R$_4$-CA & 716165 & 10616 & 32915 & 315 & $10^7$ & 625.6 & 5372039 & 1.45 & 2.46 & 1357591 & 4.99  \\ \hline\hline
		\end{tabular}
		}
	 \endgroup
\end{table*}

\begin{table*}[t]
\centering
\caption{\textbf{Performance benchmarking of the quantum-secure token exchange.} Performance metrics were measured using Hyperledger Caliper for core decentralized finance operations: Swap (asset exchange), Stake (liquidity provision via token locking), Unstake (liquidity withdrawal), and Search (ledger state and history queries). Succeeded and Failed denote the counts of processed transactions. Send Rate represents the transaction submission rate; Throughput, the rate of confirmed and committed transactions; and Average Latency, the end-to-end confirmation time (min/max in parentheses).
The benchmark was conducted on the six-node token exchange deployment described in the main text, comprising five consensus participants and one CA node. The application system was implemented in a VMware-based environment running Ubuntu~20.04 with 16~GB of RAM, and Hyperledger Caliper was used to generate and measure the transaction workloads.
For each operation type, the submitted workload was swept from 1000 to 10000 transactions to evaluate success rate, throughput saturation, and latency growth under increasing load. The zero-failure counts across all tested operations indicate that the quantum-secured ordering layer can support the tested decentralized-finance workloads without transaction loss under the evaluated conditions.
Across the four workloads, throughput increases with the submitted transaction rate at low and moderate loads and then shows operation-dependent saturation at the highest loads. In the 10000-transaction tests, Search reaches the highest throughput of 805.3~transactions per second, followed by Swap at 721.9~transactions per second, Stake at 684.9~transactions per second, and Unstake at 667.6~transactions per second. The average latency remains below one second for most 1000--5000 transaction tests and increases at heavier loads, consistent with queueing as the application approaches its measured throughput limit.}
\label{table_caliper}

\begingroup
\fontsize{10}{12}\selectfont
\setlength{\tabcolsep}{8pt}
\renewcommand{\arraystretch}{1}

\begin{tabular}{*{6}{c}}
\hline\hline
\rule[-1ex]{0pt}{3.5ex}
Operation & Succeed & Failed &
\shortstack{Send Rate\\(transactions per second)} &
Average Latency (s) &
\shortstack{Throughput\\(transactions per second)} \\
\hline
\rule[-1ex]{0pt}{3.5ex}
Swap     & 1000  & 0 & 100.8 & 0.25 (0.13--0.44) & 96.5 \\
\hline
\rule[-1ex]{0pt}{3.5ex}
Swap     & 3000  & 0 & 300.7 & 0.32 (0.14--0.97) & 276.7 \\
\hline
\rule[-1ex]{0pt}{3.5ex}
Swap     & 5000  & 0 & 500.7 & 0.49 (0.16--0.81) & 479.8 \\
\hline
\rule[-1ex]{0pt}{3.5ex}
Swap     & 8000  & 0 & 800.3 & 2.08 (0.15--3.47) & 653.4 \\
\hline
\rule[-1ex]{0pt}{3.5ex}
Swap     & 10000 & 0 & 999.9 & 3.70 (0.17--5.88) & 721.9 \\
\hline

\rule[-1ex]{0pt}{3.5ex}
Stake    & 1000  & 0 & 100.8 & 0.26 (0.12--1.06) & 91.1 \\
\hline
\rule[-1ex]{0pt}{3.5ex}
Stake    & 3000  & 0 & 300.7 & 0.32 (0.15--1.09) & 274.5 \\
\hline
\rule[-1ex]{0pt}{3.5ex}
Stake    & 5000  & 0 & 500.6 & 0.50 (0.16--1.15) & 454.3 \\
\hline
\rule[-1ex]{0pt}{3.5ex}
Stake    & 8000  & 0 & 800.3 & 2.31 (0.16--3.88) & 638.7 \\
\hline
\rule[-1ex]{0pt}{3.5ex}
Stake    & 10000 & 0 & 998.8 & 3.82 (0.17--6.00) & 684.9 \\
\hline

\rule[-1ex]{0pt}{3.5ex}
Unstake  & 1000  & 0 & 100.8 & 0.25 (0.13--0.40) & 96.9 \\
\hline
\rule[-1ex]{0pt}{3.5ex}
Unstake  & 3000  & 0 & 300.6 & 0.32 (0.14--1.11) & 273.2 \\
\hline
\rule[-1ex]{0pt}{3.5ex}
Unstake  & 5000  & 0 & 500.6 & 0.58 (0.15--1.19) & 454.0 \\
\hline
\rule[-1ex]{0pt}{3.5ex}
Unstake  & 8000  & 0 & 800.4 & 2.15 (0.15--3.36) & 648.9 \\
\hline
\rule[-1ex]{0pt}{3.5ex}
Unstake  & 10000 & 0 & 1000.6 & 4.18 (0.17--6.97) & 667.6 \\
\hline

\rule[-1ex]{0pt}{3.5ex}
Search   & 1000  & 0 & 100.8 & 0.25 (0.13--1.06) & 91.1 \\
\hline
\rule[-1ex]{0pt}{3.5ex}
Search   & 3000  & 0 & 300.7 & 0.29 (0.14--0.99) & 274.4 \\
\hline
\rule[-1ex]{0pt}{3.5ex}
Search   & 5000  & 0 & 500.6 & 0.39 (0.15--0.99) & 459.0 \\
\hline
\rule[-1ex]{0pt}{3.5ex}
Search   & 8000  & 0 & 800.6 & 1.55 (0.15--2.56) & 689.4 \\
\hline
\rule[-1ex]{0pt}{3.5ex}
Search   & 10000 & 0 & 1000.6 & 2.44 (0.16--3.99) & 805.3 \\
\hline\hline

\end{tabular}
\endgroup
\end{table*}

\subsection{Consensus Rate Calculation}
The consensus rate $R_{\mathrm{con}}$ denotes the number of successful consensus instances the system can reach per second. Since the consensus layer relies on three-party QDS operations with different participants, which consume quantum secret keys from various links, the consensus rate is limited by the link with the lowest key generation rate.

We first determine the communication complexity $C(N)$, which is the total number of three-party QDS operations required for one consensus instance. The protocol has two phases. In the order distribution phase, the master node sends orders to $N-1$ backup nodes, which requires $N-1$ QDS operations. In the circular gathering phase, each of the $N-1$ backup nodes starts a gathering cycle. Each cycle consists of $N-1$ steps. This results in $(N-1) \times (N-1)$ operations. Summing these two parts gives the total complexity as $C(N) = (N-1) + (N-1)^2 = N^2 - N$.

For a signature length of $L_{\mathrm{sig}}$, each QDS operation consumes $3 L_{\mathrm{sig}}$ bits of secret keys~\cite{2023-Yin}. Thus, the consensus rate is given by:
\begin{equation}
    R_{\mathrm{con}} = \frac{R_{\min}}{3 L_{\mathrm{sig}} \times C(N)} = \frac{R_{\min}}{3 L_{\mathrm{sig}} \times (N^2-N)}.
\end{equation}

\section{Detailed System Architecture and Data Interaction}\label{app:system}
Fig.~\ref{data_interaction} presents the architecture and data interaction workflow of the token exchange application, illustrating the coordinated operation of the frontend interface, the backend service, and the underlying blockchain network. This three-layer architecture facilitates bidirectional data flow, implementing the six-stage process described in the application schematic (Fig.~\ref{network_architechure} c) and ensuring consistency between conceptual design and system implementation.

1. Frontend Layer: A responsive Vue.js--based web application provides a comprehensive token exchange management interface. It integrates modules for dashboard monitoring, user and token administration, liquidity pool management, trading execution, and transaction history visualization. These functional modules are orchestrated by a centralized state management mechanism that maintains application state consistency across routing, lifecycle events, and data binding. User interactions, such as initiating token swaps or staking assets into liquidity pools, are translated into structured HTTP requests and transmitted to the backend through dedicated service APIs. This design enables visualization of blockchain states, including network status, account balances, and pool reserves, while maintaining a clear separation between presentation logic and system execution.

2. Backend Layer: Built on Spring Boot, the microservices architecture handles business logic and blockchain interaction. Incoming requests are validated, normalized, and processed by service components that interface with the blockchain through the Fabric SDK. For state-changing operations, the backend constructs transaction proposals, invokes chaincode on designated endorsing peers, collects endorsement responses, and submits validated transactions to the ordering service. Configuration components define critical network parameters, including organizational identities, channel settings, and network topology, ensuring consistent and secure interaction with the distributed ledger. Upon transaction completion, execution results and state updates are propagated back to the frontend for user-facing visualization and auditing.

3. Blockchain Network Layer: This layer provides distributed consensus services based on the QBA protocol. It consists of orderer nodes and endorsing peers belonging to exchange and liquidity provider organizations. Orderer nodes execute the circular QBA protocol, using quantum key distribution to achieve Byzantine fault tolerance against classical and quantum adversaries. Endorsing peers simulate chaincode and validate semantics for operations such as token issuance and automated market maker swaps. Transaction logic is enforced by smart contracts that update user balances and liquidity pool reserves according to deterministic rules, including constant-product pricing mechanisms. Once consensus is reached, ordered transactions are committed to the distributed ledger, where both immutable transaction logs and world states are maintained in logically isolated namespaces.

End-to-end data interaction follows a closed-loop workflow. First, users trigger state changes via the frontend-such as depositing assets into liquidity pools or executing token swaps-which are captured and dispatched as structured API requests. Subsequently, the backend processes these requests, constructing transaction proposals that are sent to endorsing peers. These peers simulate chaincode execution and return signed endorsements. Following endorsement, the backend submits the transaction to the ordering service, where the QBA protocol ensures global consensus on transaction order. Finally, validated blocks are committed to the ledger, updating the world state (e.g., adjusting LP token totals or user balances). This state transition is propagated back through the backend to the frontend, allowing users to visualize the finalized transaction history and updated asset balances.

\section{QKD Implementation}\label{app:qkd}
The QKD implementation combines chip-integrated quantum transmitters and receivers with finite-key secure key extraction, as summarized in Fig.~\ref{Chip_setup}.

\subsection{Chip-integrated experimental setup}

As shown in Fig.~\ref{Chip_setup}, we implement chip-integrated quantum signal transmitters and receivers at the quantum physical layer based on silicon photonic integrated circuits.

The integrated quantum signal transmitter deployed at each user node consists of a laser diode (LD) generating phase-randomized optical pulses, a silicon-based encoder chip (Encoder) dedicated to high-speed preparation of decoy states and polarization states, an off-chip variable optical attenuator (VOA) and the corresponding driving electronics (Electronics I/O).
The CA  is equipped with an integrated quantum receiver that includes a silicon-based decoder chip (Decoder) with polarization-tracking capability, multi-channel superconducting nanowire single-photon detectors (SNSPDs), a time-to-digital converter (TDC), a personal computer (PC), and the associated driving electronics.

Optical pulses emitted from the LD are coupled into the encoder chip through a one-dimensional grating coupler (1DGC).
The encoder integrates an intensity modulator (IM) and a polarization modulator (Pol-M).
The IM adopts a Mach--Zehnder interferometer (MZI) architecture composed of two multimode interferometers (MMIs), a pair of thermo-optic modulators (TOMs), and a pair of carrier-depletion modulators (CDMs).
The TOMs provide static phase biasing, while the CDMs enable high-speed phase modulation, thereby supporting fast and precise modulation of the decoy-state intensities.

After intensity modulation, the optical signal is routed via on-chip waveguides to the polarization modulator (Pol-M).
The Pol-M consists of an MZI driven by a pair of CDMs, cascaded with another pair of CDMs, and finally connected to a two-dimensional grating coupler (2DGC).
By driving the two pairs of CDMs, path encoding signal is implemented on chip and is subsequently mapped to polarization encoding signal by the 2DGC.
Based on this mechanism, the four polarization states required by the BB84 protocol are generated, with the quantum state expressed as
\[
\left| \psi \right\rangle = \frac{1}{\sqrt{2}}\left( \left| H \right\rangle + e^{\mathbf{i}\theta}\left| V \right\rangle \right),\quad
\theta \in \left\{0, \frac{\pi}{2}, \pi, \frac{3\pi}{2}\right\},
\]
where $\theta \in \{0, \pi\}$ ($\theta \in \{\pi/2, 3\pi/2\}$) correspond to the $Z$ ($X$) basis states.
The optical signal is subsequently attenuated to the single-photon level by the off-chip VOA and launched into the quantum channel.

At the receiver side, the optical signal arriving from the quantum channel is first coupled into the silicon-based decoder chip through an on-chip spot-size converter (SSC).
The polarization-encoded signal is then converted into on-chip path-encoded information by a polarization splitter-rotator (PSR) and further passively distributed to the $Z$- or $X$-basis measurement paths by two symmetric MMIs.
 Polarization state analysis is performed by two on-chip polarization controllers, each consisting of a pair of TOMs and an MZI driven by another pair of TOMs. By precisely tuning the TOM driving voltages, projective measurements in the $Z$ and $X$ bases are realized. The measured optical pulses are routed via SSCs to multi-channel SNSPDs for detection, and the experimental data are acquired and post-processed using the TDC and PC.

Both the silicon-based encoder and decoder chips are fabricated using a standard silicon photonics process at a commercial foundry and are packaged with high reliability to ensure long-term stable operation.
The encoder chip has a footprint of $6 \times 3~\mathrm{mm^2}$ and is hermetically sealed in a butterfly package with an overall volume of $20 \times 11 \times 5~\mathrm{mm^3}$.
The decoder chip has a footprint of $1.6 \times 1.7~\mathrm{mm^2}$ and is assembled using board-level integration, resulting in a total volume of $3.95 \times 2.19 \times 0.90~\mathrm{cm^3}$.
The CDMs in the encoder chip are driven by an arbitrary waveform generator to enable high-speed quantum state encoding, while all TOMs are precisely controlled by programmable linear DC power supplies.

\subsection{Device characterization}
The integrated quantum devices are systematically characterized at a repetition rate of 50~MHz using optical pulses with a central wavelength of 1549.17~nm and a pulse width of 200~ps. The intensity modulator on the encoder chip exhibits a dynamic extinction ratio of approximately 18~dB, which satisfies the intensity modulation requirements of the signal and decoy states in the one-decoy-state BB84 protocol. The total insertion loss of the decoder chip is approximately 7.9~dB, and the $Z$-basis selection probability is maintained at approximately 50$\%$. The SNSPDs are uniformly calibrated to achieve a detection efficiency of 60$\%$.

Within this configuration, four polarization states $|H\rangle$, $|V\rangle$, $|D\rangle$, and $|A\rangle$ are prepared using the on-chip polarization modulator and subsequently measured by the receiver chip via projective detection. The average polarization extinction ratio measured for the four states is approximately 23~dB. These results demonstrate that the implemented polarization encoder and decoder chips enable high-fidelity state preparation and measurement, thereby effectively suppressing the quantum bit error rate (QBER) and establishing a robust device-level foundation for stable system operation and reliable protocol performance.

\subsection{Secure Key Generation and Finite-Key Analysis}
Under optimized system parameters, $10^{7}$ valid $Z$-basis detection events are cumulatively collected on each quantum link to form the raw key, which is subsequently processed through classical post-processing to extract the final secure key. The secure key extraction procedure comprises two main stages. First, information reconciliation is performed to correct errors originating from the quantum channel and device imperfections, ensuring that all user nodes share an identical raw key with the certification authority (CA). Next, under finite-key conditions, parameter estimation and security analysis are conducted to evaluate the achievable secure key rate, after which privacy amplification is applied to the reconciled key to generate the final secure key bits.

Following Ref.~\cite{2018-Rusca}, the upper bound of the secure key rate under finite-key conditions is expressed as
\begin{equation}\label{SKR_finite}
\begin{split}
R \le \Biggl\{ & s_{z}^{(L,0)} + s_{z}^{(L,1)} \Bigl[ 1 - h(\phi_{z}^{U}) \Bigr] \\
& - \mathrm{leak}_{\mathrm{EC}}  - 6 \log_2 \Bigl( \frac{19}{\varepsilon_{\mathrm{sec}}} \Bigr)
 - \log_2 \Bigl( \frac{2}{\varepsilon_{\mathrm{cor}}} \Bigr) \Biggr\}/t,
\end{split}
\end{equation}
where $t$ denotes the duration of a single data-acquisition period; $s_{z}^{(L,0)}$ and $s_{z}^{(L,1)}$ are the lower bounds of the effective vacuum and single-photon events in the $Z$ basis, respectively; $\phi_{z}^{U}$ is the upper bound on the phase-error rate for single-photon events in the $Z$ basis; $\mathrm{leak}_{\mathrm{EC}}$ represents the information leakage during error correction; $\varepsilon_{\mathrm{sec}}$ and $\varepsilon_{\mathrm{cor}}$ are the secrecy and correctness parameters, respectively; $h(x)$ is the binary Shannon entropy function.

\section{Detailed Experimental Results}\label{app:exp-results}
Table~\ref{table_rawdate} shows the detailed experimental results.

\section{Detailed Caliper Performance Test Results}\label{app:caliper}
Table~\ref{table_caliper} shows the detailed Caliper performance test results for the quantum-secure token exchange application.

\bibliography{ref}

\end{document}